\documentclass[conference]{IEEEtran}
\IEEEoverridecommandlockouts
\usepackage{cite}
\usepackage{amsmath,amssymb,amsfonts}
\usepackage{algorithmic}
\usepackage{graphicx}
\usepackage{subcaption}
\usepackage{textcomp}
\usepackage{xcolor}
\usepackage{wrapfig}
\usepackage{caption}
\usepackage{colortbl}
\usepackage{footnote}
\usepackage{gensymb}
\usepackage{breqn}
\usepackage{mathtools}
\usepackage{xfrac}
\def\BibTeX{{\rm B\kern-.05em{\sc i\kern-.025em b}\kern-.08em
    T\kern-.1667em\lower.7ex\hbox{E}\kern-.125emX}}
\begin{document}

% INCLUDE THE WORD - EQUIPMENT
% \title{Design and Implementation of a PUF-based Hardware Security Solution for IEDs
\title{PUF Probe: A PUF-based Hardware Authentication Equipment for IEDs 
% {\footnotesize \textsuperscript{*}Note: Sub-titles are not captured in Xplore and should not be used}
\thanks{This work was carried out under PGCoE, funded by Power Grid Corporation of India.}
}
%\maketitle
\author{\IEEEauthorblockN{Vishal D. Jadhav}
\IEEEauthorblockA{\textit{Dept. of Electronic Systems Engg.} \\
\textit{Indian Institute of Science}\\
Bangalore, India \\
vishal294j@gmail.com}
\and
\IEEEauthorblockN{Narahari N. Moudhgalya}
\IEEEauthorblockA{\textit{Dept. of Electronic Systems Engg.} \\
\textit{Indian Institute of Science}\\
Bangalore, India \\
n.narahari.m@gmail.com}
\and
\IEEEauthorblockN{Tapabrata Sen}
\IEEEauthorblockA{\textit{Dept. of Electronic Systems Engg.} \\
\textit{Indian Institute of Science}\\
Bangalore, India \\
sen.tapu@gmail.com}
\and
\IEEEauthorblockN{T V Prabhakar}
\IEEEauthorblockA{\textit{Dept. of Electronic Systems Engg.} \\
\textit{Indian Institute of Science}\\
Bangalore, India \\
tvprabs@iisc.ac.in}
}
\maketitle

% \and
% \and
% \and
% \and
% \and

\begin{abstract}
Intelligent Electronic Devices (IEDs) are vital components in modern electrical substations, collectively responsible for monitoring electrical parameters and performing protective functions. As a result, ensuring the integrity of IEDs is an essential criteria. While standards like IEC 61850 and IEC 60870-5-104 establish cyber-security protocols for secure information exchange in IED-based power systems, the physical integrity of IEDs is often overlooked, leading to a rise in counterfeit and tainted electronic products. This paper proposes a physical unclonable function (PUF)-based device (IED\textsubscript{PUF} probe) capable of extracting unique hardware signatures from commercial IEDs. These signatures can serve as identifiers, facilitating the authentication and protection of IEDs against counterfeiting. The paper presents the complete hardware architecture of the IED\textsubscript{PUF} probe, along with algorithms for signature extraction and authentication. The process involves the central computer system (CCS) initiating IED authentication requests by sending random challenges to the IED\textsubscript{PUF} probe. Based on the challenges, the IED\textsubscript{PUF} probe generates responses, which are then verified by the CCS to authenticate the IED. Additionally, a two-way authentication technique is employed to ensure that only verified requests are granted access for signature extraction. Experimental results confirm the efficacy of the proposed IED\textsubscript{PUF} probe. The results demonstrate its ability to provide real-time responses possessing randomness while uniquely identifying the IED under investigation. The proposed IED\textsubscript{PUF} probe offers a simple, cost-effective, accurate solution with minimal storage requirements, enhancing the authenticity and integrity of IEDs within electrical substations.
\end{abstract}


\section{Introduction}\label{one}
% In the electronic power industry, the intelligent electronic device (IED) is an integrated microprocessor based controller of power system equipment \cite{Hussain}. IEDs use digital controls and include relays, fault recorders, circuit breakers etc. Thus a typical IED configuration may consist of analog/digital inputs from power equipment and sensors, ADC and DAC, DSP, internal memory, Flex-logic and display \cite{Hussain}. Several of these IEDs can work collaboratively, exchanging data over a network, to facilitate data collection and control action. However, this also makes them susceptible to attacks. The attack vectors are classified as: (i) Hardware related, (ii) Firmware related, (iii) Software related, and (iv) communication related \cite{JWang}.
% Through these attacks, the cryptographic keys could be stolen, rogue nodes could be introduced in the network. Thus, compromising the entire IED-based power systems. 

In the field of electric power industry, the intelligent electronic device (IED) serves as an integrated microprocessor-based controller for power system equipment \cite{Hussain}. IEDs utilise digital controls and includes relays, fault recorders, circuit breakers, and other components. Consequently, a typical IED configuration consists of analog \& digital inputs from power equipment and sensors, analog-to-digital (ADC) modules, Digital Signal Processors (DSP), internal memory, Flex-logic, and a display \cite{Hussain}. These IEDs operate collaboratively, exchanging data over a network to facilitate data collection and control actions. However, this interconnectedness also renders them susceptible to various types of attacks, classified as hardware-related, firmware-related, software-related, and communication-related vectors \cite{JWang}. Such attacks can result in the theft of cryptographic keys or introduction of rogue nodes, and thus compromising the overall security of IED-based power systems.

% Moreover, the risks of counterfeiting and tainting is a serious concern with all types of embedded electronic devices. The risks associated with counterfeiting and tainting are introduction of malware, unauthorized, or scrap parts. Hence, the irrefutable identity of the IED is at the heart of securing the IED-based power system automation. Most existing solutions that are primarily based on digital code stored in non-volatile memory and are prone to identity-theft attacks. 

PUFs leverage inherent manufacturing variations to create identities that are extremely difficult to replicate. These identities serve as a foundation for various applications, such as device identification, generating encryption keys for data storage and transmission, and ensuring traceability to prevent the introduction of clones or rogue devices into the network \cite{Hamlet}. PUFs can be employed either as standalone identities or to complement and reinforce the identities generated through stored identifiers.

In this work, PUFs are utilised to generate identities for commercial IEDs. The primary contributions of this paper are summarized as in the following. 
\begin{itemize}
    \item The paper proposes an PUF for IED ($\mathrm{IED_{PUF}}$) probe that extracts hardware signatures in real-time from COTS components and hardware modules of a commercial IED. 
    \item The proposed $\mathrm{IED_{PUF}}$ probe employs a two-way authentication protocol with the central computer system in the process of verifying the authenticity of the IED.
    \item A simple, efficient algorithm has been proposed to largely reduce storage overhead present in traditional solutions.
    %feature
    % \item The developed PUF probe is simple, low-cost, portable, and accurately identify IED devices in real-time.
    %feature
    % \item Unlike conventional solutions, the proposed PUF probe offers minimal storage requirement due to an efficient algorithm.
    \item The extracted signatures from different IEDs possess randomness for random challenges, but repeatable (for same challenge) and uniquely identifies the IED.
    % \item The signature is extracted from COTS components present in the IED. Thus, the solution exploits the variations in the different hardware modules of different IEDs.
    %\item The two-way authentication algorithm ensures that only authorized personnel will get access for signature extraction.
\end{itemize}

The features of the proposed $\mathrm{IED_{PUF}}$ probe can be summarized as: (i) low-cost, portable, accurate solution for authenticating IEDs in real-time; (ii) offers minimal storage requirement; (iii) the solution exploits the variations in the different hardware modules of the IEDs; (iv) the solution is IED device-specific.

This paper is organised as follows: Section \ref{two} discusses the working methodology of the proposed $\mathrm{IED_{PUF}}$ probe, Section \ref{three} presents the details of the $\mathrm{IED_{PUF}}$ probe electronics, experimental results are reported in Section  \ref{four}, and Section \ref{five} concludes the paper.

\section{Related Work}
To address these challenges, in this work Physically Unclonable Functions (PUFs) \cite{Joshi} are employed to generate unique digital fingerprints or identifiers for hardware devices.
PUFs can be generated from custom circuits \cite{JLee,GSuh}, off-the-shelf components \cite{DunCan,Vaidya1}, and memories \cite{Holcomb}. Key properties of PUFs include uniqueness, randomness, reliability, and repeatability \cite{Joshi}.

Moreover, the risks of counterfeiting and tainting is a serious concern with all types of embedded electronic devices \cite{IEC20243}. These risks encompass the infiltration of malware and the unauthorised usage of substandard or counterfeit components. Hence, the irrefutable identity of the IED is at the heart of securing the IED-based power system automation. Existing solutions primarily rely on digital codes stored in non-volatile memory, which are susceptible to identity theft attacks.

{While the literature extensively covers the extraction of PUF signatures from electronic systems, limited research has been conducted on extracting PUF signatures from commercially available information and communication technology (ICT) or Internet of Things (IoT) products. \cite{Vaidya1} proposes a solution to generate IoT device identifier by extracting PUF signatures from clock oscillators and ADCs in-built in MCUs. Machine-learning (ML) model has been employed for creating the device identifiers using extracted PUF signatures. In similar works in \cite{Vaidya2, Vaidya3}, signatures are extracted from sources such as the current profile of electromagnetic relays and MCU General-Purpose Input/Output (GPIO) pins. In other works in \cite{Jason,Edwards}, unique identifiers are generated for printed circuit boards (PCB) by leveraging variations in surface mount passive components and wire trace patterns. A arbiter PUF-based hardware security solution for protecting smart power grids from cyber threats is proposed in \cite{Mohammad}. This solution introduces a new algorithm that utilizes PUF signatures and timestamps from a GPS clock. %An application of PUF in cryptography key (such as AES key) generation. 
These works discuss PUF signature extraction and its usage in creating identifiers for IoT devices. However, the scopes for PUF extraction directly from an existing COTS IoT product (such as a commercial IED) have not been explored largely. Moreover, conventional PUF solutions typically require significant memory resources to store PUF challenge-response pairs.}

% Physically Unclonable Functions (PUFs) \cite{Joshi} have been used to generate identity or digital fingerprint for hardware devices. Since they leverage the manufacturing variations, the identity is generated based on innate property of the device; making it almost impossible to replicate the identity. This identity could then be used as a primitive for different applications like: (i)Identification - To ensure that devices over the network are identified undisputedly, (ii)Cryptography - To generate key for encryption during storage and transmission of data and (iii)Traceability -- To avoid any clones/rogues getting introduced into the network. PUFs could be generated from custom circuits \cite{JLee,GSuh} and even from off-the-shelf components \cite{DunCan,Vaidya}, and memories \cite{Holcomb}. Some of the properties of PUF are -- Uniqueness, Randomness, Reliability, and Repeatability.

% There is large literature on extraction of PUF signature from electronic systems. However, fewer works have focused on extracting PUF signatures from commercial-off-the-shelf (COTS) information and communication technology (ICT) products. 

\section{Hardware Architecture of $\mathrm{IED_{PUF}}$ probe}\label{two}
The detailed architecture of the $\mathrm{IED_{PUF}}$ probe is brought out in this section. First, the different sources for PUF extraction from a commercial IED (SEL-311C \cite{SEL}) are illustrated. Following this, the complete system-level schematic of the $\mathrm{IED_{PUF}}$ probe is presented. It needs to be clarified here that the proposed $\mathrm{IED_{PUF}}$ probe is optimized for the IED SEL-311C. IEDs from different manufacturers may require modification of the $\mathrm{IED_{PUF}}$ probe electronics. However, the communication protocol between $\mathrm{IED_{PUF}}$ probe and CCS, signature formation and authentication process does not require any modification. As a result, developing customized PUF probes for different IED manufacturers can be considered as a reasonable solution.

\subsection{Measurement Principle for PUF Extraction from Diode Cut-in region, internal regulated voltages, and time period}\label{twoA}
% In this section, we go over the experimental setup for extracting signatures from IED. We also go over how two-way authentication is implemented between servers and IEDs.
A typical IED has several output relay ports, communication modules (include serial and ethernet), data conversion modules (include ADC and DAC), power Supply modules (include LDOs, DC-DC converters), current \& potential transformers, GPS, digital \& analog ports, etc. Thus, identifying the possible sources for PUF signature extraction from the SEL-311C IED was carried out at first. %The output relay ports of the IED and internal analog signals that can be extracted through serial terminal of the IED were explored.
The identified PUF sources are -- (i) Cut-in region voltages of the protective diodes across the output relay ports, (ii) internal regulated voltages, and (iii) internal clock period.

\subsubsection{Diode Voltages}\label{twoA1}
%This section discusses  extraction of signatures and hardware implementation from output ports of an IED. 
% \textcolor{red}{Output relay ports of an IED are often used for isolation of circuits in the grid.} 
IEDs utilize relay outputs in power substations for critical functions such as tripping circuit breakers, activating alarms, controlling substation devices, enforcing interlocking rules, and enabling emergency shutdowns.
On inspection of the SEL-311C IED, it was found that the output relay ports have mechanical relays and associated fast recovery protection diodes connected across it. 
%Further inspection brought out that these relays have protection diodes connected in parallel to them. 
% The IED output ports contains mechanical relays. As a protective measure, diodes are connected across each relay port.   
Due to fabrication process variations, the characteristics of these diodes, such as voltage-current response, are expected to differ. Table \ref{Diode_volts} presents the recorded voltages obtained by measuring the diode voltages using a $5\sfrac{1}{2}$-digit multimeter. The table shows the variation among the diodes voltages at the output ports of an IED, with twelve adjacent ports chosen for this demonstration. %The following diode voltages were obtained by passing approximately 1~mA current in forward bias condition, through each of the twelve protection diode, sequentially. 
\begin{table}[t!]
%\centering
\caption{Voltage variations among the output port diodes of an IED}
\label{Diode_volts}
\begin{tabular}{|c|c|c|c|}
\hline
\textbf{\begin{tabular}[c]{@{}c@{}}Output \\ Port No.\textsuperscript{*}\end{tabular}} &
  \textbf{\begin{tabular}[c]{@{}c@{}}Diode \\ Voltage (mV)\end{tabular}} &
  \textbf{\begin{tabular}[c]{@{}c@{}}Output \\ Port No.\end{tabular}} &
  \textbf{\begin{tabular}[c]{@{}c@{}}Diode \\ Voltage (mV)\end{tabular}} \\ \hline
OUT201 & 500.8 & OUT207 & 514.3 \\ \hline
OUT202 & 516.8 & OUT208 & 502.0 \\ \hline
OUT203 & 513.4 & OUT209 & 516.1 \\ \hline
OUT204 & 514.2 & OUT210 & 517.6 \\ \hline
OUT205 & 511.5 & OUT211 & 515.1 \\ \hline
OUT206 & 514.1 & OUT212 & 514.4 \\ \hline
\end{tabular}
\newline
%\flushleft
\footnotesize \rule{0pt}{2.0ex}Notes:\quad \textsuperscript{*} Same as in the SEL-311C IED
\end{table}

\begin{figure}[b!]
\centerline{\includegraphics[width=1.0\linewidth]{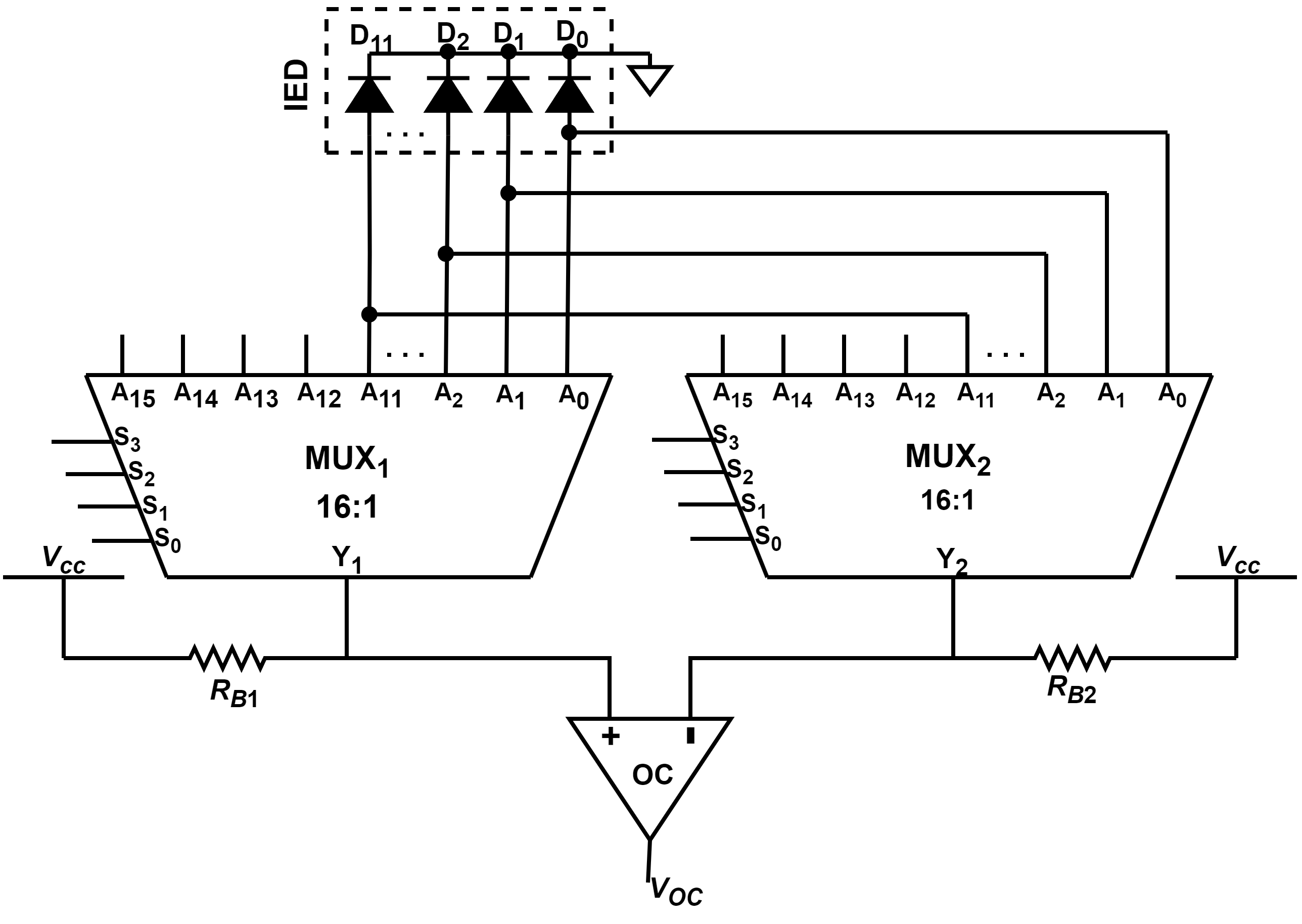}}
\caption{Basic schematic for PUF extraction from IED output protection diodes}
\label{basic_ckt}
\end{figure}

\subsubsection{Signature Extraction from Diode Cut-in Region}\label{twoA2}
%Comments : R1 is equal to R2
%SIG can be represented in HEX
%In terms of application, the primary objective of the developed system is creating signatures using the identified sources of PUFs.
Our approach considers the diode cut-in voltage  region and associated combinations of diode pairs. %One possible approach can be comparing the forward voltage drops across the diodes, considering all possible combinations of diode pairs. Fig. \ref{basic_ckt} shows the basic implementation.
As mentioned earlier, twelve output relay ports, and thus, corresponding twelve diodes ($\mathrm{D_0}$, $\mathrm{D_1}$, ..., $\mathrm{D_{11}}$) were selected for measurement. $\mathrm{MUX_1}$ and $\mathrm{MUX_2}$ (${16}\times{1}$ multiplexers) operate as analog multiplexer/demultiplexer. The diodes were connected to the terminals $\mathrm{A_0}$, $\mathrm{A_1}$, ..., $\mathrm{A_{11}}$ of $\mathrm{MUX_1}$ and $\mathrm{MUX_2}$ as in Fig. \ref{basic_ckt}. The terminals $\mathrm{Y_1}$ and $\mathrm{Y_2}$ of $\mathrm{MUX_1}$ and $\mathrm{MUX_2}$ were pulled to the supply voltage $V_{cc}$ using the resistors $R_{B1}$ and $R_{B2}$. Under this configuration, when one of the terminals (say, $A_0$) of $\mathrm{MUX_1}$ is connected to $\mathrm{Y_1}$, (through appropriate logic levels at the multiplexer selector pins), a finite current (say, $I_S$) flows  to forward bias the diode $\mathrm{D_0}$. 
Similarly, using $\mathrm{MUX_2}$, one of the other diodes (between $\mathrm{D_1}$ to $\mathrm{D_{11}}$) can be forward biased. %by passing an current equal to $I_S$, sourced through $V_{cc}$, the resistor $R_B$, and the output $\mathrm{Y_2}$, as in Fig. \ref{basic_ckt}. 
The terminals $\mathrm{Y_1}$ and $\mathrm{Y_2}$ were directly connected to a low offset  voltage ($<1$~mV) differential comparator chip (OC) as shown in Fig. \ref{basic_ckt}.

Under the aforementioned setup, the twelve diode voltages were successively compared to those of the other eleven diodes. As a result, we will get $N(N-1)$ diode comparisons (where $N=12$ in this case). A 1-bit output will be produced for every comparison. Let us consider one instance to elaborate this.
The terminals $A_0$ of $\mathrm{MUX_1}$ and $A_1$ of $\mathrm{MUX_2}$ were selected %by applying suitable voltage levels to $\mathrm{S_{0}}$- $\mathrm{S_{3}}$ 
for comparing the cut-in region voltages of a diode-pair, say $\mathrm{D_0}$ and $\mathrm{D_1}$. 
Under this setting, the voltages at {Y\textsubscript{1}} and {Y\textsubscript{2}} terminals are, ideally, equal to the voltages (say, $V_{D0}$ and $V_{D1}$) across $\mathrm{D_0}$ and $\mathrm{D_1}$. The comparator output $V_{OC}$ can be expressed as:
\[{V_{OC}} = {1}\,\,{\rm{for}}\,\,{V_{D0}} > {V_{D1}}\,\,\,\,{\rm{and}}\,\,\,\,{V_{OC}} = 0\,\,{\rm{for}}\,\,{V_{D0}}\leq {V_{D1}}\]
%will be high or low based on the magnitudes of $V_{D1}$ and $V_{D2}$. Thus, $V_{AC}$ is essentially equivalent one bit of signature. 

The procedure was repeated and $V_{D0}$ was compared to the diode voltages $V_{D1}$ to $V_{D_{11}}$ to generate a 11-bit signature. Repeating the above procedure, each diode was compared with the remaining diodes. Thus, considering twelve output ports of the IED, ${12}\times{11}$ or 132-bit signature was formed. The recorded bit-streams (say, $SIG_1$ and $SIG_2$) for two such IEDs are shown below. The signatures are expected to be unique to the individual IEDs under investigation.
\begin{flalign*}
SIG_1 = 000000000000101111111001100010010000101011\\01
0000100000010000101010010000101111010000100000000\\000101111110001111111111001111111111101101111110000
%\label{eq1}
\end{flalign*}
\begin{multline*}
SIG_2 = 000010100001101011111111100010100001111011\\11
1111000000000000101010111101000010000001101010100\\001101010110001101010111001101011111101000010000000
%\label{eq1}
\end{multline*}

% Resistor-Diode voltage divider has been made as shown in the figure. We chose voltage-supply and resistance value such that it will keep diodes in forward bias region. After measuring current and voltages with various resistor values, 3k ohm resistor found out to be the better option for biasing at a supply voltage of 3.3V. This will allow a small amount of current around 0.8-0.9mA to pass through the diode.
\subsubsection{IED Regulated Voltages and Internal Oscillator Clock Period}\label{twoA3}
%Over any type of communication not only serial communication
%Repeatability and uniqness can be portrayed here

{The SEL-311C IED offers a serial port API to extract and display the clock period (say, $T_P$) and internal regulator voltages (say, $V_{RG}$) of the IED.} %range of configurable commands and internal status access points through serial communication. 
%Few of the commands are to access the status of the internal components. For eg.“TIM” command is used to get status or to set IED time(24-hour time). The PUF probe is connected to the IED using serial to TTL converter and it can communicate with IED. 
%{Upon investigating these commands, it was found that analog signals, such as internal regulated voltages (say, $V_{RG}$) and clock period (say, $T_P$) of the IED can be obtained through serial communication interface.} 
%For the operation of digital circuits and microprocessor DC voltage is needed. As IED are installed in power grids, the main source of power for IEDs will be the AC voltage from the grid. For IEDs operations, this AC voltage will be stepped down using internal transformers. This voltage is then converted to DC voltage for digital operations. 
{The IED SEL-311C contains three internal voltage regulators with nominal voltages of 3.3~V, 5~V, and 15~V. However, actual voltages may exhibit slight deviations due to fabrication process variations across regulator chips. As a result, $V_{RG}$ can serve as potential PUF sources for different IEDs.} 
%In SEL311C these regulated voltages are accessible over command. Using PUF Probe we can access these internal regulated voltages over serial communication using the “STA” command. After repeated measurements of these voltages it is found out to be unique and repeatable signatures and is strong puf. 
{Furthermore, the time period $T_P$ of the internal oscillator within the IED was extracted through serial communication using appropriate commands. The nominal value of $T_P$ is 20~nS. The deviation of $T_P$ is anticipated to be distinct for each IED, and thus presenting an additional parameter suitable for PUF generation.}

% {In order to affirm the uniqueness of $V_{RG}$ and $T_P$, two SEL-311C IEDs were tested. 
% Table \ref{VR_TP} shows the results for this comparison. The $V_{RG}$ and $T_P$ values were recorded over serial terminal a few hundred times to check repeatability, further affirming their viability as distinctive characteristics for PUF-based applications.}
\begin{table}[h!]
\centering
\caption{Variations in $V_{RG}$ and $T_P$ values in two IEDs}
\label{VR_TP}
\begin{tabular}{|c|c|c|c|c|}
\hline
\textbf{\begin{tabular}[c]{@{}c@{}}IED\\ No.\end{tabular}} & \textbf{\begin{tabular}[c]{@{}c@{}}$\bf{V_{RG1}}$ \\ = 3.3 V\end{tabular}} & \textbf{\begin{tabular}[c]{@{}c@{}}$\bf{V_{RG2}}$ \\= 5 V\end{tabular}} & \textbf{\begin{tabular}[c]{@{}c@{}}$\bf{V_{RG3}}$ \\= 15 V\end{tabular}} & \textbf{\begin{tabular}[c]{@{}c@{}}$\bf{T_P}$ =\\ 20 nS\end{tabular}} \\ \hline
IED-1 & 3.26 & 4.95 & 14.85 & 19.999945 \\ \hline
IED-2 & 3.29 & 4.99 & 14.86 & 20.000063 \\ \hline
\end{tabular}
\end{table}

\subsubsection{Signature Extraction from Regulated Voltages and Clock Period}\label{twoA4}

{In order to affirm the uniqueness of the regulated three voltages ($V_{RG1}$, $V_{RG2}$, and $V_{RG3}$) and $T_P$, two SEL-311C IEDs were tested. 
Table \ref{VR_TP} shows the results for this comparison. The regulated voltages and $T_P$ values were recorded over serial terminal a few hundred times to check the repeatability, and thus affirming their viability as distinctive characteristics for PUF-based applications.}
%Thus, ∆TP of 55 fS and -63 fS are represented as 0110111 and 1111111 => this can be made fixed size for eg 8 bits 16 bits in 2's complement
% first prototype also can be kept here
{Table \ref{VR_TP} reveals that the uniqueness in the regulated voltages and $T_P$ values for the two IEDs can be observed in their deviations (say, $\Delta{V_{RG}}$ and $\Delta{T_{P}}$) from the respective nominal values. 
For example, in the case of IED-1,  $\Delta{V_{RG1}}=40$~mV, $\Delta{V_{RG2}}=50$~mV, and $\Delta{V_{RG3}}=150$~mV. %values are 40~mV, 50~mV, and 150~mV for the three regulated voltages. 
On the other hand, for IED-2, the corresponding values for $\Delta{V_{RG1}}$, $\Delta{V_{RG2}}$, and $\Delta{V_{RG3}}$ are 10~mV, 10~mV, and 140~mV, respectively. Similarly, $\Delta{T_{P}}$ values for the two IEDs are measured as 55~femtoseconds (fs) and $-63$~fs, respectively. %Therefore, these $\Delta{V_{RG}}$ and $\Delta{T_{P}}$ values can be regarded as unique signatures for the IEDs. 
Signatures were formed using these parameters by converting the absolute decimal values of $\Delta{V_{RG1}}$, $\Delta{V_{RG2}}$, $\Delta{V_{RG3}}$, and $\Delta{T_{P}}$ to corresponding binary (disregarding order information such as units (e.g., mV, fs) and any trailing zeros in the $\Delta{V_{RG}}$ values). Additionally, an extra bit was added to the left of each binary code (as the most significant bit (MSB)) to denote polarity, with a 1 indicating a negative value and 0 representing a positive value. For instance, $\Delta{T_{P}}$ values of 55~fS and $-63$~fS were transformed into binary as 0110111 and 1111111, respectively. Similarly, the $\Delta{V_{RG}}$ values were also converted to binary, and finally, all the four parameters were concatenated in the same order as presented in Table \ref{VR_TP}. The resulting 22-bit signatures for the two IEDs are illustrated below.}
\begin{equation*}
SIG_{1} = 0010000101011110110111
%\label{eq1}
\end{equation*}
\begin{equation*}
SIG_{2} = 0000100001011101111111
%\label{eq1}
\end{equation*}

% \subsubsection{Internal Oscillator Clock Period}\label{twoA3}
% Every microcontroller based system has internal clock source for eg. Crystal oscillator or RC oscillator. Similarly, IEDs will have oscillators. Due to variations in the fabrication process these oscillators will give slightly different clock periods. Like regulator voltages, the clock period of a SEL311C can be accessed over serial communication using commands. According to the datasheet of a SEL311C  the clock period of an internal oscillator can be accessed using command “TIM Q”. After repeated measurements it is found out to be a strong signature and is repeatable and unique. It is found to be constant up to 6 decimal places. Comparison of clock period of oscillator from different IEDs has been shown in the following table. 
\subsection{Working Circuit for $\mathrm{IED_{PUF}}$ probe}\label{twoB}

\begin{figure}[t!]
\centerline{\includegraphics[width=1.0\linewidth]{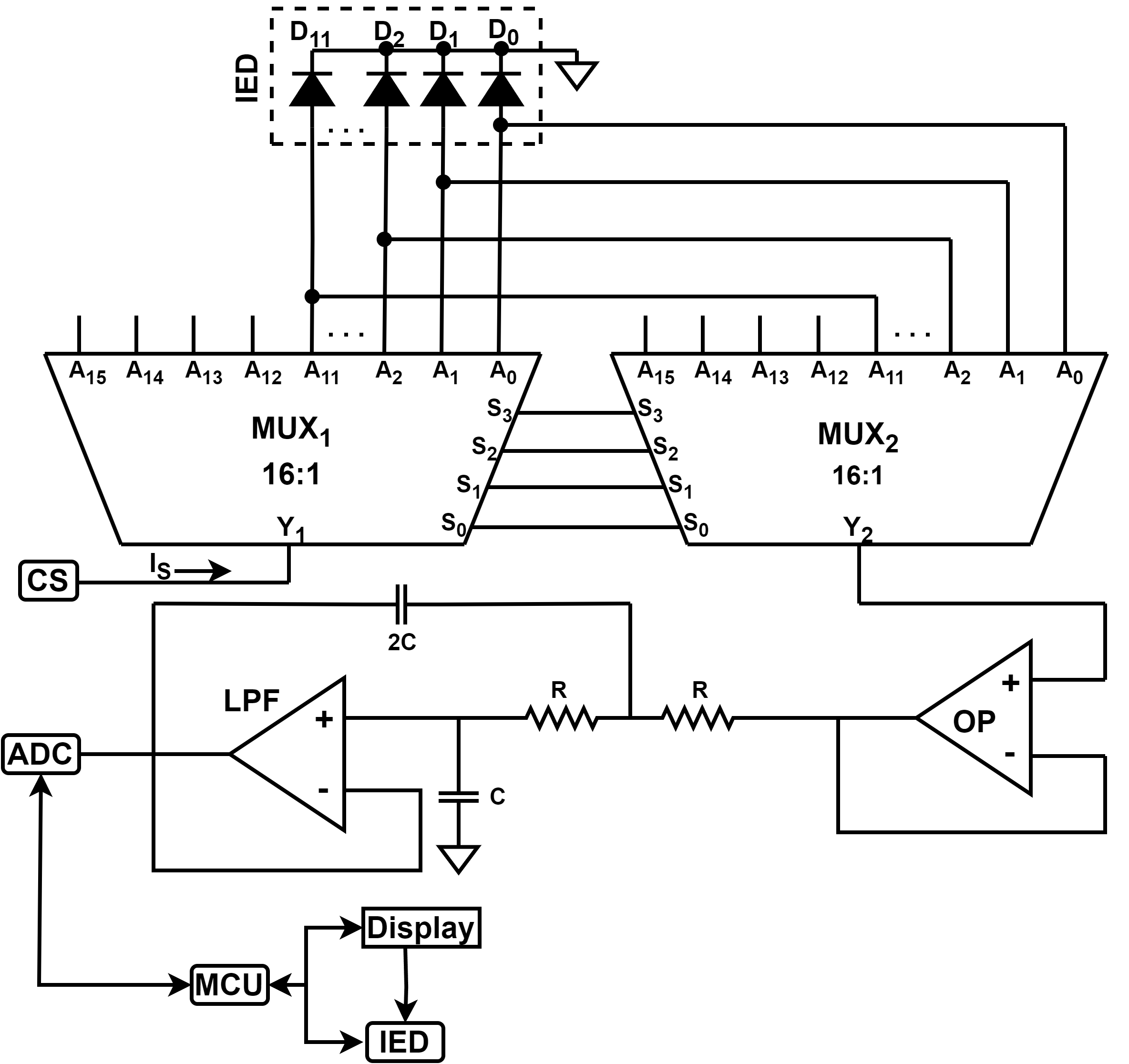}}
\caption{System-level schematic of $\mathrm{IED_{PUF}}$ probe}
\label{improve_ckt}
\end{figure}

%\subsubsection{Improved Circuit Implementation}
Earlier, in Section \ref{twoA2}, a basic circuit schematic for extracting signatures from the twelve output port diodes ($\mathrm{D_0}$, $\mathrm{D_1}$, ..., $\mathrm{D_{11}}$) of a SEL-311C IED was presented. The measurement principle was simple. In the working circuit we address issues, such as high on-resistance (say, $R_{ON}$) of the multiplexer switches, offset voltage of comparator, mismatches in the resistors $R_{B1}$ and $R_{B2}$, variation of $R_{ON}$ with power-supply and temperature changes. 
%As a result, it was imperative to improve the design for reliable signature extraction. 

{The system-level schematic of the working $\mathrm{IED_{PUF}}$ probe is shown in Fig. \ref{improve_ckt}. It incorporates various electronic hardware modules, including analog multiplexer/demultiplexer ($\mathrm{MUX_1}$ and $\mathrm{MUX_2}$), an ADC, a microcontroller unit (MCU), a constant current source (CS), and a display unit.  Similar to Fig. \ref{basic_ckt}, twelve diodes ($\mathrm{D_0}$, $\mathrm{D_1}$, ..., $\mathrm{D_{11}}$) are connected to the inputs ($\mathrm{A_0}$, $\mathrm{A_1}$, ...) of $\mathrm{MUX_1}$. %These diodes are essentially the protective diodes connected across the output ports of the IED. 
In this design, the diodes are forward biased by a constant current (say, $I_S$) generated by the CS. The CS unit utilizes a precision reference voltage and a PNP transistor-based configuration \cite{lm385}. %$I_S$ is passed through each diode, sequentially, by selecting appropriate inputs of $\mathrm{MUX_1}$ through the selector pins $\mathrm{S_0}$ to $\mathrm{S_3}$. 
As shown in Fig. \ref{improve_ckt}, $I_S$ flows from the $\mathrm{Y_1}$ terminal of $\mathrm{MUX_1}$ to the terminals $\mathrm{A_0}$, $\mathrm{A_1}$ through $\mathrm{A_{11}}$, sequentially. The selection of $\mathrm{MUX_1}$ channels is controlled by applying appropriate logic levels to the selector pins $\mathrm{S_{0}}$ to $\mathrm{S_{3}}$.} %of a DEMUX is pulled to the Vcc using a Resistor R1. All output ports of an IED were connected to the output terminal of the DEMUX such that when any one port is selected it will allow small amount of current to flow from Vcc to the input terminal of DEMUX and then finally passing through output. This will keep the diode in forward biasing mode. After that Voltage across diode has taken as a signature.

{The voltage (say, $V_{O1}$) at the $\mathrm{Y_1}$ terminal of $\mathrm{MUX_1}$ is related to the voltages across the diodes. However, $V_{O1}$ also contains a component related with the voltage drop ($V_{sw}$) across the internal resistance ($R_{ON}$) of $\mathrm{MUX_1}$ caused by the current $I_S$ flowing through $\mathrm{MUX_1}$. The $V_{sw}$ component is significantly high and depends on temperature and power supply variations. To address this issue, $\mathrm{MUX_2}$ has been introduced. The selector pins of $\mathrm{MUX_1}$ and $\mathrm{MUX_2}$ are tied together, as in Fig. \ref{improve_ckt}, ensuring the selection of the same channels in both $\mathrm{MUX_1}$ and $\mathrm{MUX_2}$ during channel sweeping. The terminal $\mathrm{Y_2}$ of $\mathrm{MUX_2}$ is connected to an op-amp (OP)-based voltage follower circuit. By leveraging the low bias current of the op-amp, the effect of $R_{ON}$ is compensated.} 

{The output of the op-amp is then passed through a simple second-order low-pass filter before being fed into a high-resolution ADC. The cutoff frequency of the filter is set at approximately 15~Hz. %This way, the signal input to the ADC is band-limited from higher frequency. 
From Table \ref{Diode_volts}, it can be observed that the differences among the diode voltages can be of the order of 1~mV or even lower. As a result, selection of a suitable ADC is crucial for accurately resolving these voltages. Moreover, the measured diode voltages are static in nature. Considering these, a DC optimized Delta-Sigma ($\Delta$-$\Sigma$) ADC is selected as will be detailed in Section \ref{fourA}.} %This way, the requirement for the comparator present in Fig. \ref{basic_ckt}, could be eliminated.}

{As mentioned in Section \ref{twoA3}, the regulated voltages and $T_P$ parameters of the IED were acquired through serial terminal. The deployed MCU performs this serial communication operation with the IED and is also responsible for controlling other operations of the $\mathrm{IED_{PUF}}$ probe, such as applying appropriate logic levels to the multiplexer selector pins, managing the ADC \& the display modules, implementing algorithms for signature extraction, and transmission operations.}

\begin{figure}[b!]
\centerline{\includegraphics[width=1.0\linewidth]{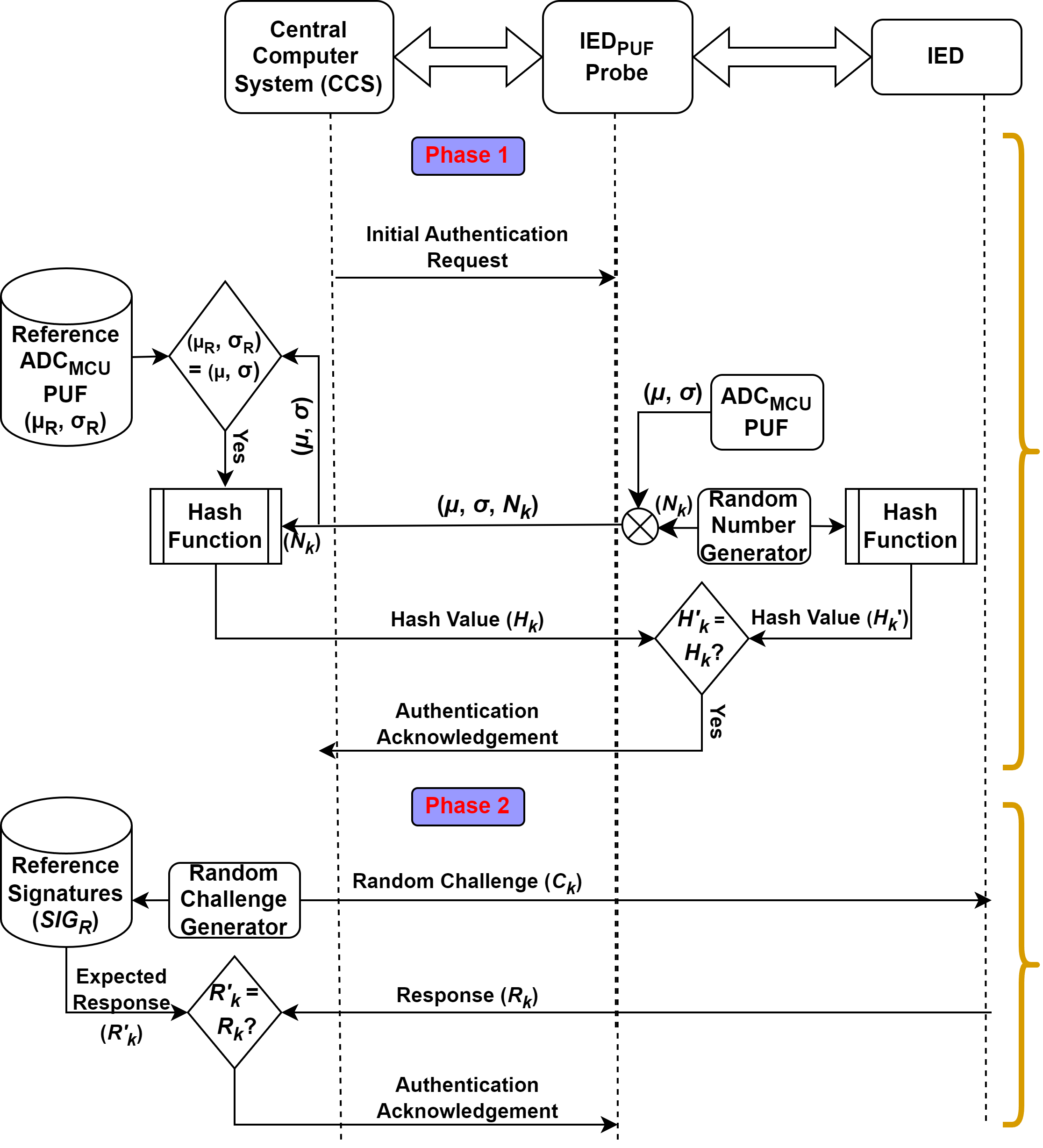}}
\caption{Complete flow-chart showcasing the IED authentication mechanism performed by the CCS and the $\mathrm{IED_{PUF}}$ probe}
\label{auth_flow}
\end{figure}

\section{Working Principle of $\mathrm{IED_{PUF}}$ probe}\label{three}
Fig. \ref{auth_flow} depicts the authentication process for $\mathrm{IED_{PUF}}$ probe, as well as, the IED device. 
The signature extraction process involves a series of sequences executed when the CCS requests for IED signature extraction. The following steps outline this procedure:
\begin{enumerate}
    \item The CCS initiates communication by pinging the $\mathrm{IED_{PUF}}$ probe through the serial terminal. 
    \item The $\mathrm{IED_{PUF}}$ probe and CCS execute Phase I, which involves verifying the authentication of the IED access request originating from the CCS, as well as, authenticating the $\mathrm{IED_{PUF}}$ probe. 
    \item Upon successful authentication in Phase I, the CCS proceeds to Phase II by transmitting random port numbers (corresponding to the twelve output port under investigation) as challenges to the $\mathrm{IED_{PUF}}$ probe.
    \item The $\mathrm{IED_{PUF}}$ probe generates a response to the challenges presented by the CCS in Phase II. This response is then transmitted back to the CCS, where it undergoes further verification processes to authenticate the IED.
\end{enumerate}
%The CCS securely stores in its database a 154-bit reference signature (say, $SIG_R$) for individual IEDs present in the electrical substation. %The first 132 bits of $SIG_R$, starting from the MSB, corresponds to the signature extracted from the IED output port protection diodes, while the remaining 22 bits represent signatures extracted from $V_{RG}$ and $T_P$. 
Phase I performs mutual authentication between the CCS and the $\mathrm{IED_{PUF}}$ probe. To accomplish this, we extract ADC PUF (ADC\textsubscript{MCU}) hardware signature using the procedure elaborated in \cite{Vaidya1}. Using this procedure, we estimate the mean ($\mu$) and standard deviation ($\sigma$) of the 12-bit ADC present in the MCU. Phase II focuses on implementing a challenge-response-based mechanism to authenticate the IED on the CCS side. 
%In Phase I, mutual authentication between the CCS and the $\mathrm{IED_{PUF}}$ probe is executed. %the request made by the CCS to extract hardware signatures from the IED. 
%Phase II performs IED authentication utilizing $SIG_R$. 
These two phases will be further elaborated in the subsequent subsections. 

% In PUF challenge-response pair(CRPs)-based authentication, it is assumed that the challenge is coming from ‘authorised’ Central Computer System(CCS). So anyone with an IED's puf CRPs can pretend to be CCS and can access IED. To tackle this problem, PUF probe will extract puf response of an IED only if the CCS is authorised by PUF Probe. Detailed description of a 'Two Way Authentication' has been explained in this section. Man-in-the-middle attack is out of the scope of this paper.

% \begin{figure}[h]
% \includegraphics[width=8cm]{}
% \end{figure}

\subsection{Phase I - Authenticating CCS}\label{threeA}
% In Phase I of authentication, the signature extraction request issued by the CCS is verified. This phase also authenticates that the signature is extracted using a legitimate $\mathrm{IED_{PUF}}$ probe. For this purpose, the variations in the in-built ADC (ADC\textsubscript{MCU}) of the MCU have been leveraged to extract PUF signatures for the $\mathrm{IED_{PUF}}$ probe \cite{Vaidya1}. The ADC\textsubscript{MCU} has been configured in differential mode. Fig. \ref{ADC1} and \ref{ADC2} show that the histogram of ADC output values for two identical MCU devices. The mean (say, $\mu$) and standard deviation (say, $\sigma$) values are indicated in the figures. Notably, these values demonstrate distinct relative differences between the MCU devices. $\mu$ and $\sigma$ values serve as the $\mathrm{IED_{PUF}}$ probe signatures. Phase I executes in the following sequence. 

The Phase I starts with an initial authentication request issued by the CCS and ends with an authentication acknowledgement. %initiates the authentication process by sending an initial authentication request to the $\mathrm{IED_{PUF}}$ probe. 
Upon receiving the authentication request, the $\mathrm{IED_{PUF}}$ probe sends three parameters to the CCS: a randomly generated number $N_k$, $\mu$, and $\sigma$. These two ADC parameters are obtained from the ADC\textsubscript{MCU} PUF. %over 10,000 samples obtained by shorting specific ADC inputs.
The CCS compares these values with previously stored references. %, allowing for a tolerance of $\pm{5}\%$ error. 
If the comparison yields a match, the authenticity of the $\mathrm{IED_{PUF}}$ probe is validated. Following this stage, the CCS transmits a hash value $H_k$, generated using a MD5 hash function and $N_k$ as an input. %is processed. $N_k$ serves as an input for the MD5 hash function which is used by both the CCS and the genuine $\mathrm{IED_{PUF}}$ probe. The CCS generates a hash value, $H_k$, at its end and transmits it back to the $\mathrm{IED_{PUF}}$ probe. 
The $\mathrm{IED_{PUF}}$ probe also generates its hash response $H_k'$ to the same input $N_k$. $H_k$ and $H_k'$ are compared for a match to  authenticate the CCS access request. %by the $\mathrm{IED_{PUF}}$ probe. If the two hash values match, the CCS access request is successfully authenticated. 
%Consequently, the CCS and the $\mathrm{IED_{PUF}}$ probe proceed with the PUF signature extraction process in Phase II. 
Conversely, if $H_k$ and $H_k'$ do not match, the Phase I is deemed unsuccessful, and access is denied. The aforementioned sequences are illustrated under Phase I in Fig. \ref{auth_flow}.

\begin{figure}[t!] 
    \centering
  \subfloat[\label{ADC1}]{%
      \includegraphics[width=0.49\linewidth]{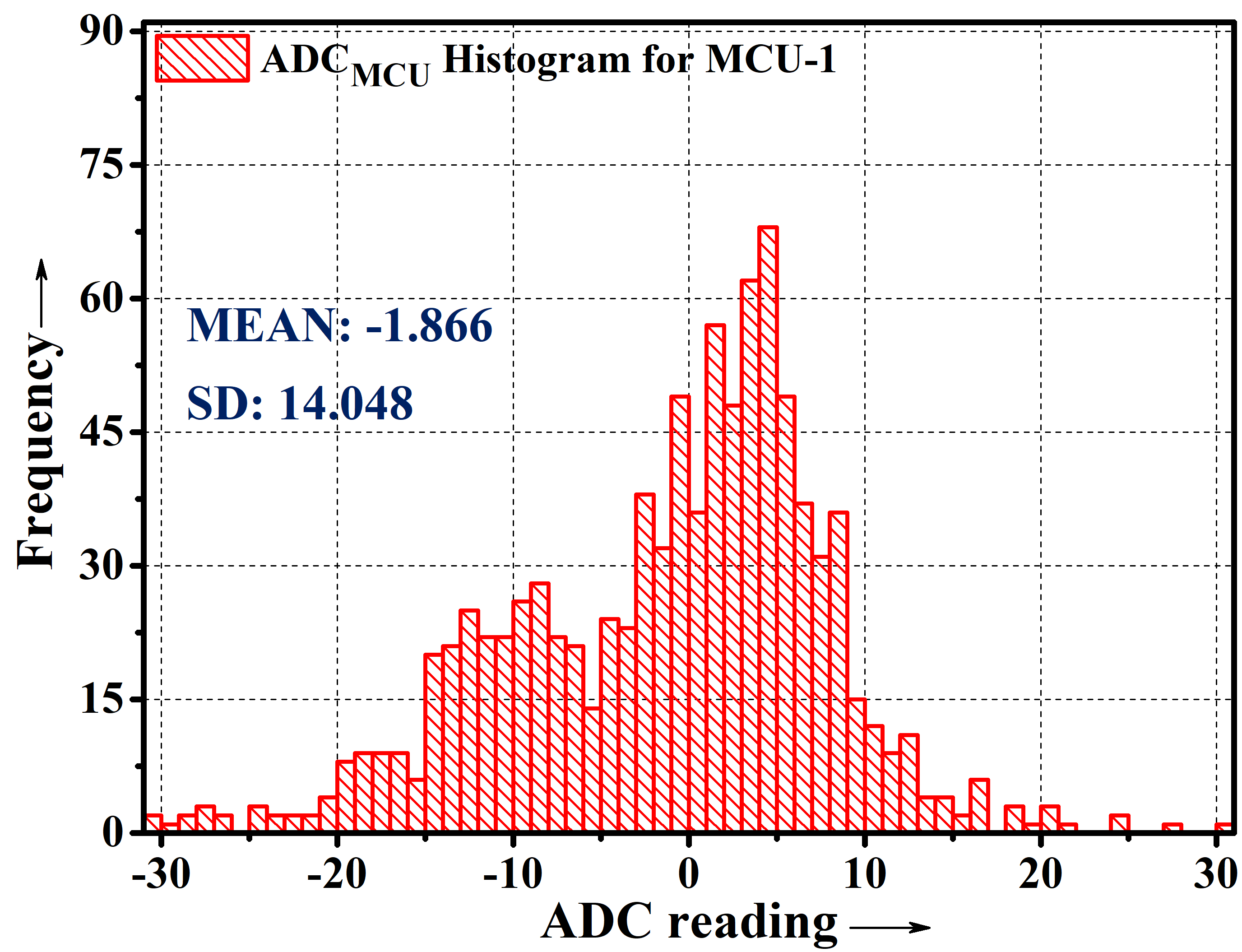}}
    \hfill
  \subfloat[\label{ADC2}]{%
        \includegraphics[width=0.49\linewidth]{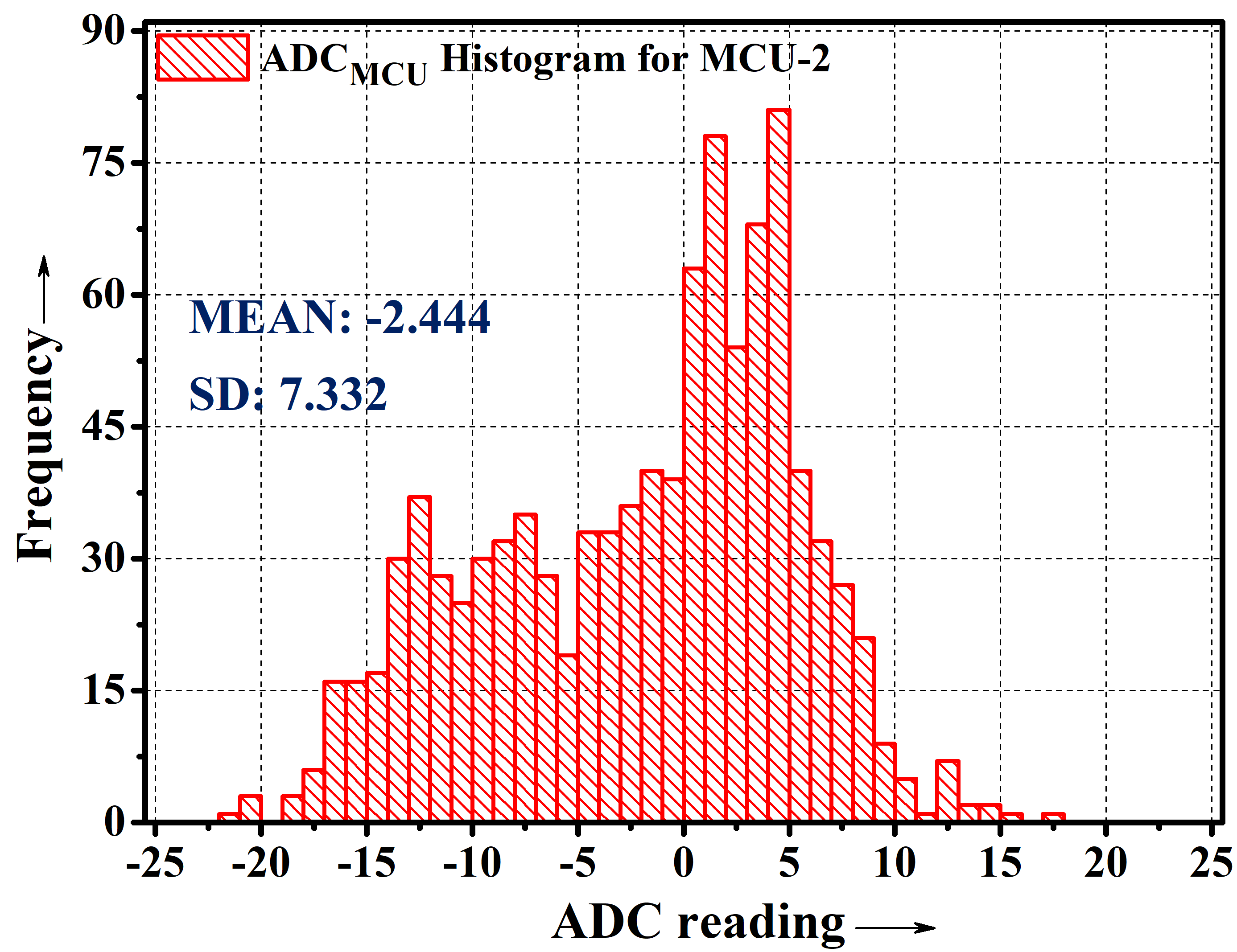}}
  \caption{Histograms of ADC output values for two identical MCUs acquired in differential mode .}
  \label{ADCdiff} 
\end{figure}

%For completeness ADC\textsubscript{PUF} results are shown in Fig. \ref{ADC1} and \ref{ADC2}. 
Figs. \ref{ADC1} and \ref{ADC2} show that the histogram of ADC output values for two identical MCU devices. The measurements were taken over 10,000 samples in differential mode. %$\mu$ and  $\sigma$ values are indicated in the figures.
$\mu$ and  $\sigma$ values, indicated in the figures, demonstrate uniqueness for the two MCU devices. %$\mu$ and $\sigma$ values serve as the $\mathrm{IED_{PUF}}$ probe signatures.

% \begin{figure}[!htb]
% \centerline{\includegraphics[width=1.0\linewidth]{images/Phase1.png}}
% \caption{Phase1}
% \label{fig1}
% \end{figure}

\subsection{Phase II - Authenticating IED}\label{threeB}

Upon successful completion of Phase I, the CCS proceeds with the authentication of the IED as shown under Phase II of Fig. \ref{auth_flow}. As mentioned previously, the CCS securely stores reference signatures ($SIG_R$) of all IEDs in its database. %$SIG_R$ encompasses three key components as discussed earlier.
For the authentication process, the CCS transmits a random challenge ($C_k$) to the $\mathrm{IED_{PUF}}$ probe, consisting of random port numbers (say, $N_i$ where $N\in[1,12]$ for $i=1,2,...12$). The 
different $N_i$ values corresponds to a port number, between Port 1 to Port 12.  The number of ports challenged in each $C_k$ will vary randomly, such as challenging eight ports in one instance and all twelve ports in another. After issuing the challenge, the CCS will generate an anticipated response ($R_k'$) from its database of reference signatures, which includes the following relevant data of the corresponding IED device: (i) clock period ($T_P$), (ii) three regulated voltages ($V_{RG1}$, $V_{RG2}$, and $V_{RG3}$), and (iii) a partial bit-stream, formed by utilizing the known $C_k$ and the reference $SIG_R$.
Upon receiving $C_k$, the $\mathrm{IED_{PUF}}$ probe will retrieve the signatures from the IED. Subsequently, it will generate a bit-stream specific to the challenged ports and combine it with the $T_P$ and $V_{RG}$ signatures, forming the response $R_k$ which is transmitted back to the CCS. 
The CCS compares $R_k$ with $R_k'$ to authenticate the IED. 
%If $T_P$ and $R_G$ portion of $R_k$ with $R_k'$ are identical, the matching percentage of bits in the bit-streams will be output as the authentication accuracy.  Otherwise the IED is not authenticated.
\begin{figure}[b!]
\centerline{\includegraphics[width=0.9\linewidth]{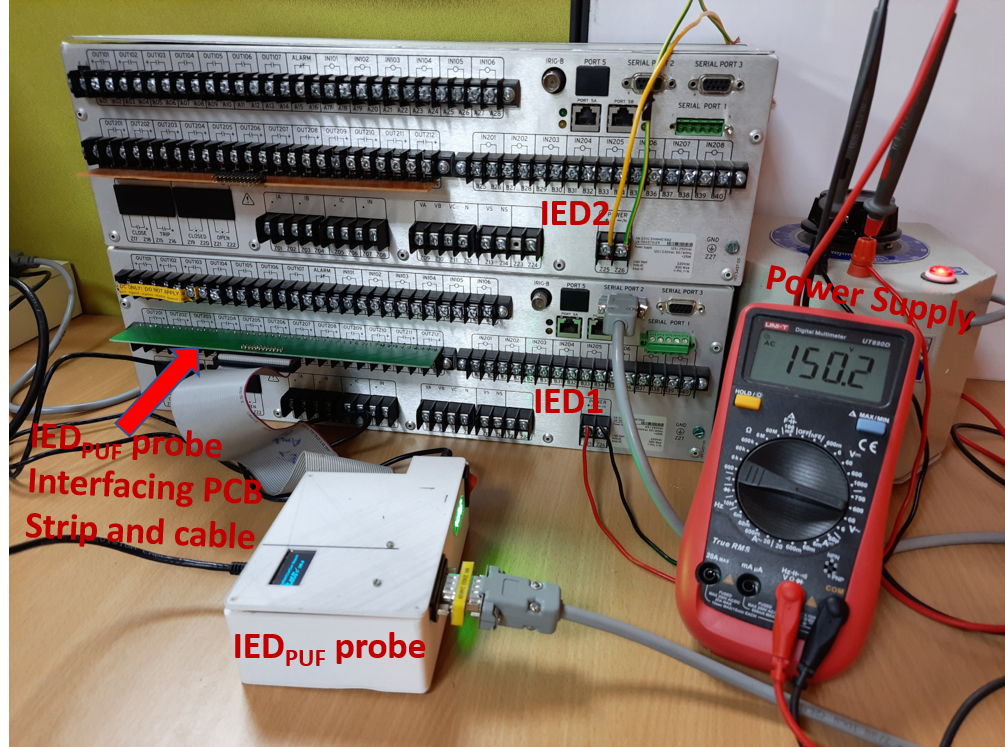}}
\caption{Experimental Setup}
\label{setup}
\end{figure}

% {In summary, the signature extraction process involves a series of sequences executed when the CCS requests for IED signature extraction. The following steps outline this procedure:}
% \begin{enumerate}
%     \item The CCS initiates communication by pinging the $\mathrm{IED_{PUF}}$ probe through the serial terminal. 
%     \item The $\mathrm{IED_{PUF}}$ probe and CCS execute Phase I, which involves verifying the authentication of the IED access request originating from the CCS, as well as, authenticating the $\mathrm{IED_{PUF}}$ probe. 
%     \item Upon successful authentication in Phase I, the CCS proceeds to Phase II by transmitting random port numbers (corresponding to the twelve output port under investigation) as challenges to the $\mathrm{IED_{PUF}}$ probe.
%     \item The $\mathrm{IED_{PUF}}$ probe generates a response to the challenges presented by the CCS in Phase II. This response is then transmitted back to the CCS, where it undergoes further verification processes to authenticate the IED.
% \end{enumerate}

\section{Experimental Evaluation}\label{four}
\subsection{Experimental Setup}\label{fourA}

{Fig. \ref{setup} shows the complete experimental setup employed for the characterization and authentication of an SEL-311C IED using the proposed $\mathrm{IED_{PUF}}$ probe. %Voltage greater than 60~V DC was supplied for powering the IED.  
%The $\mathrm{IED_{PUF}}$ probe is housed within a 3D-printed casing, effectively encapsulating all the components within the enclosure. 
As previously mentioned, twelve output ports of the IED are chosen for signature extraction, utilizing the protection diodes and a dedicated PCB strip facilitates this process. To establish the connection between the $\mathrm{IED_{PUF}}$ probe and the IED, a flat-ribbon cable and the aforementioned PCB strip are employed. The MCU module is implemented using an NRF52840-based microcontroller unit. $\mathrm{MUX_1}$ and $\mathrm{MUX_2}$ are designed using CD4067-based analog multiplexers. Furthermore, the ADC module has been implemented utilizing the ADS1115 16-bit $\Delta$-$\Sigma$ converter. The serial communication between the $\mathrm{IED_{PUF}}$ probe and the IEDs are realized through an RS232 to TTL serial interface module. The CCS is implemented on a laptop computer, and the algorithms on the CCS side are coded in Python.} 

\subsection{Experimental Results}\label{fourB}

\begin{figure}[t!] 
    \centering
  \subfloat[\label{IED1basic}]{%
      \includegraphics[width=0.49\linewidth]{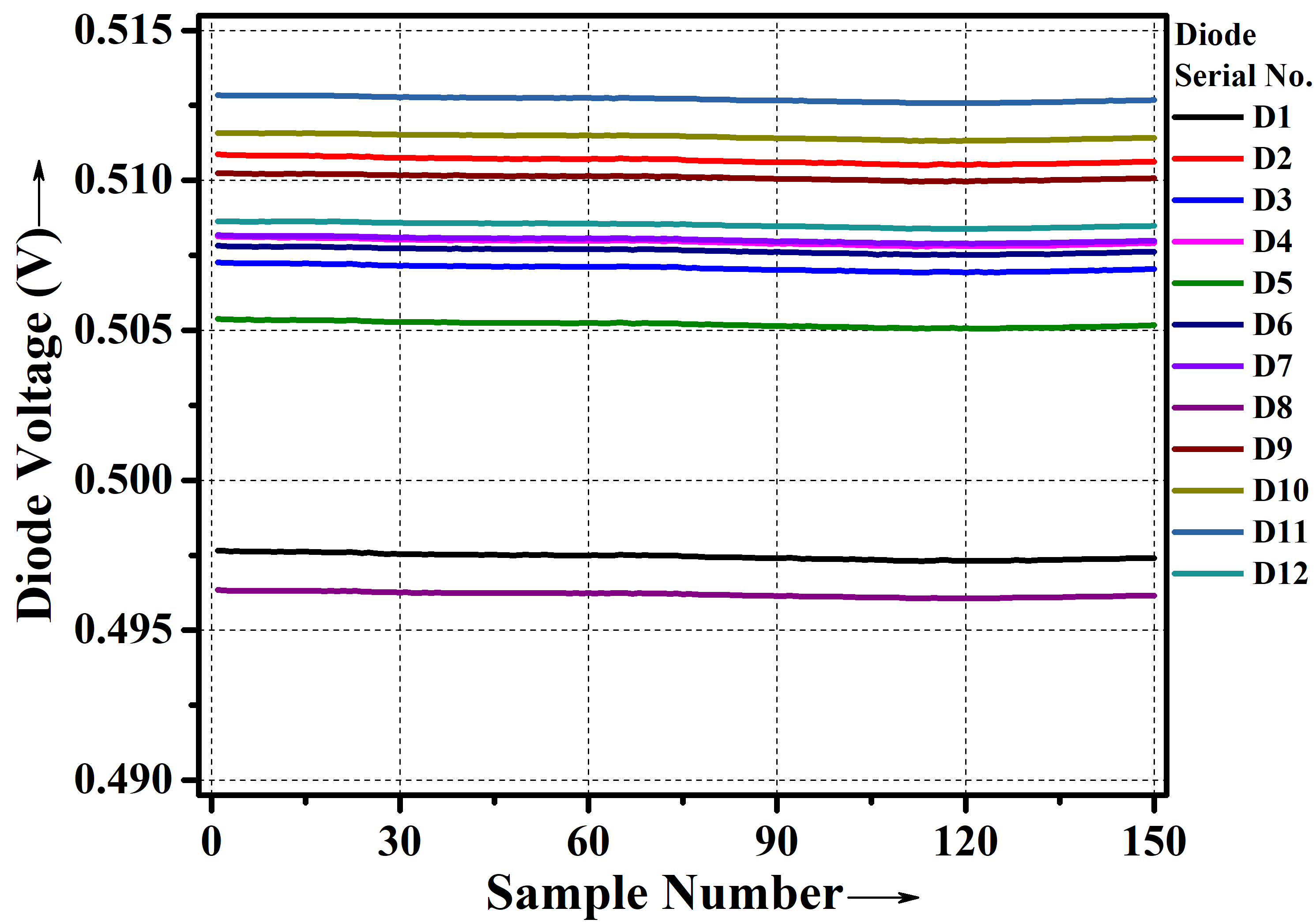}}
    \hfill
  \subfloat[\label{IED2basic}]{%
        \includegraphics[width=0.49\linewidth]{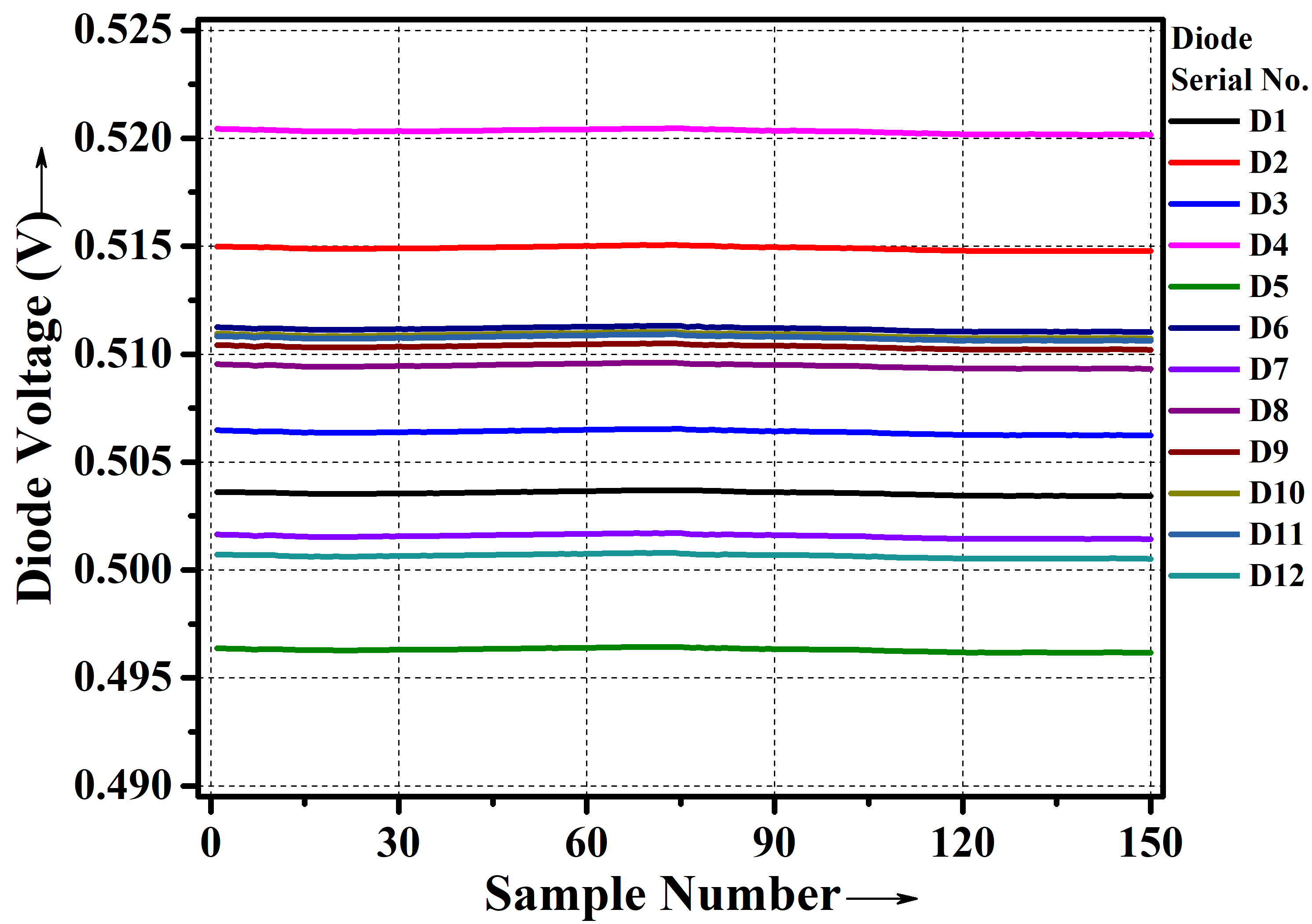}}
  \caption{Preliminary results showing the diode voltages plotted over 150 sample points. (a) response for IED-1, (b) response for IED-2.The results demonstrated repeatability and uniqueness in the measurement.}
  \label{IEDbasic} 
\end{figure}

The experiments were conducted with two commercial SEL-311C IEDs, as previously mentioned. %The following results presented in this section were derived after extensive experimentation with these IEDs. 
Preliminary results for these IEDs are illustrated in Fig. \ref{IEDbasic}. These figures display the diode voltages obtained from continuous acquisition of 150 samples at a constant current of ${1}$~mA. The twelve diode voltages of IED-1, shown in Fig. \ref{IED1basic}, exhibit variations within the range 495~mV to 513~mV. On the other hand, IED-2 demonstrates diode voltage variations ranging from 495~mV to 520~mV. It can be observed that in Fig. \ref{IED1basic}, 11 of the 12 line-plots are distinctly visible, while in Fig. \ref{IED2basic}, 10 lines are distinct. This indicates the presence of collisions in very few of the diode voltages for both IEDs. Moreover, the horizontal characteristics of the line-plots affirms excellent repeatability in the measurements. %From the color-coded line plots in Fig. \ref{IED1basic} and \ref{IED2basic}, it can be inferred that the diode voltages generate distinct patterns in both IEDs. 

\begin{figure}[t!] 
    \centering
  \subfloat[\label{IED1voltage}]{%
      \includegraphics[width=0.49\linewidth]{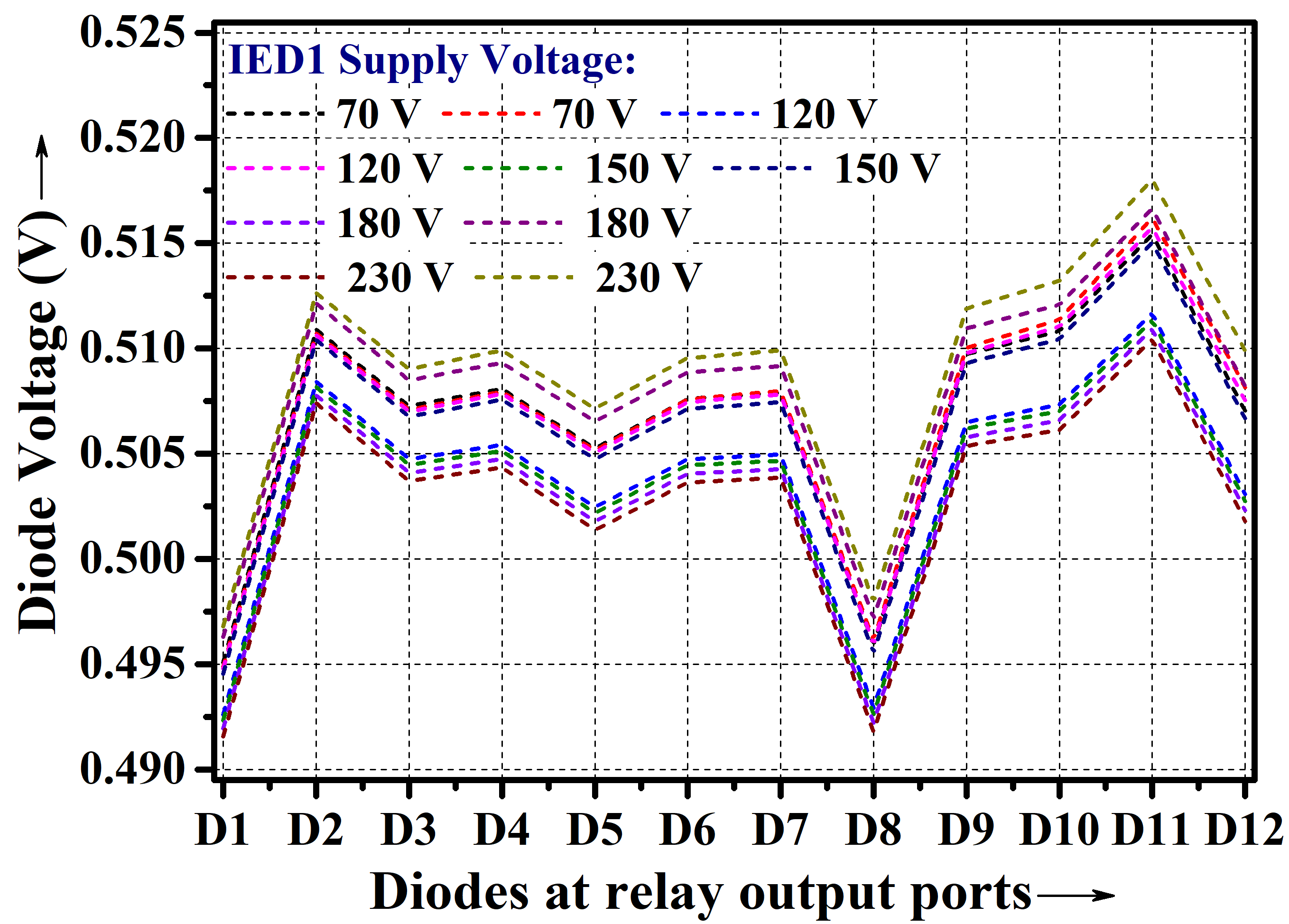}}
    \hfill
  \subfloat[\label{IED2voltage}]{%
        \includegraphics[width=0.49\linewidth]{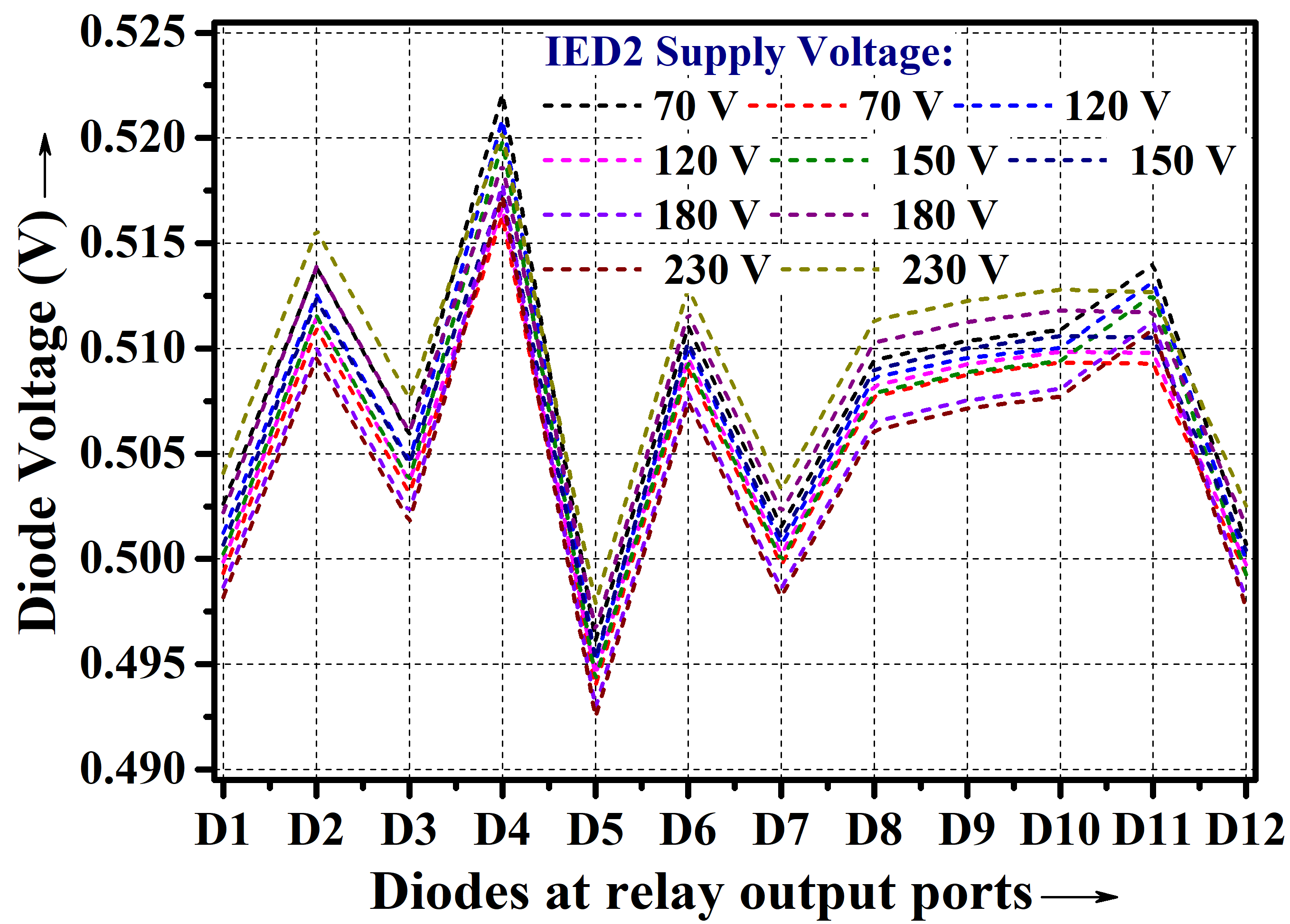}}
  \caption{Voltage characteristics of the IED output port protection diodes, taken over two weeks under varying IED supply voltage. (a) response for IED-1, (b) response for IED-2.}
  \label{IEDvoltages} 
\end{figure}

Fig. \ref{IEDvoltages} further clarifies the diode signature patterns for the two IEDs. %Detailed experimental results will be presented in Section \ref{fourB}.
The results in Fig. \ref{IEDvoltages} were obtained by monitoring the diode voltages over a span of two weeks. During this experiment, the power supply of the IEDs was varied over 70~V to 230~V to study the robustness of the signatures. Fig. \ref{IED1voltage} and \ref{IED2voltage} present the worst-case variations observed in the diode voltages for the two IEDs. It is evident from these plots that the characteristics of the two IEDs exhibit distinctive patterns, signifying their uniqueness. 

In order to examine the robustness performance of the $\mathrm{IED_{PUF}}$, the following common metrics \cite{Jason} of PUF signatures are assessed. These parameters are evaluated using the full 154-bit PUF signature derived from the datasets of Fig. \ref{IEDvoltages}. %consisting of 132 bits derived from the comparison of forward voltages of the protection diode, and the remaining 22 bits obtained from the $V_{RG}$ and $T_P$ values.

\begin{enumerate}
    \item Uniqueness: This parameter is determined by assessing the inter-device variation. This involves performing an XOR operation on the bit-strings obtained from two separate IEDs, summing the result, and then the sum is divided by the length of the bit-strings. The desired value for inter-device variation is 0.5. The worst-case values of uniqueness between IED-1 and IED-2 were evaluated to be approximately 0.31. 
    \item Stability: Intra-device variation quantifies this parameter. To evaluate stability XOR operation is performed on two bit-strings measured from a particular IED, the result is summed and the sum is divided by the length of the bit-string. Ideally, the intra-device variation is expected to be zero. The worst-case values of stability for IED-1 and IED-2 were found to be 0.09 and 0.06, respectively.
    \item Bias: Bias is calculated by dividing the sum of the bit-string by the length of the bit-string. The desired bias value is 0.5. The bias for the signatures of IED-1 and IED-2 were determined to be 0.51 and 0.52, respectively.
\end{enumerate}

\begin{table}[h!]
\centering
\caption{Occurrence of the twelve diodes in random sequence as challenge for PUF response generation}
\label{random}
\resizebox{\columnwidth}{!}{%
\begin{tabular}{|c|c|c|c|c|c|c|c|c|c|c|c|c|}
\hline
 & \textbf{S1} & \textbf{S2} & \textbf{S3} & \textbf{S4} & \textbf{S5} & \textbf{S6} & \textbf{S7} & \textbf{S8} & \textbf{S9} & \textbf{S10} & \textbf{S11} & \textbf{S12} \\ \hline
\textbf{D0} & 843 & 825 & 830 & 828 & 844 & 820 & 859 & 798 & 830 & 864 & 817 & 842 \\ \hline
\textbf{D1} & 800 & 843 & 835 & 861 & 836 & 861 & 830 & 841 & 865 & 826 & 806 & 796 \\ \hline
\textbf{D2} & 824 & 801 & 864 & 873 & 772 & 816 & 880 & 814 & 818 & 848 & 836 & 854 \\ \hline
\textbf{D3} & 836 & 847 & 794 & 855 & 793 & 783 & 886 & 853 & 836 & 830 & 798 & 889 \\ \hline
\textbf{D4} & 868 & 810 & 823 & 820 & 839 & 859 & 785 & 863 & 817 & 800 & 850 & 866 \\ \hline
\textbf{D5} & 828 & 821 & 849 & 843 & 859 & 844 & 825 & 848 & 812 & 822 & 832 & 817 \\ \hline
\textbf{D6} & 854 & 851 & 831 & 788 & 842 & 795 & 824 & 847 & 824 & 859 & 857 & 828 \\ \hline
\textbf{D7} & 851 & 836 & 827 & 811 & 887 & 848 & 817 & 848 & 788 & 840 & 826 & 821 \\ \hline
\textbf{D8} & 794 & 851 & 855 & 836 & 823 & 854 & 784 & 831 & 869 & 814 & 853 & 836 \\ \hline
\textbf{D9} & 811 & 902 & 795 & 835 & 823 & 853 & 852 & 823 & 845 & 804 & 824 & 833 \\ \hline
\textbf{D10} & 851 & 769 & 870 & 836 & 863 & 812 & 837 & 834 & 788 & 854 & 872 & 814 \\ \hline
\textbf{D11} & 840 & 844 & 827 & 814 & 819 & 855 & 821 & 800 & 908 & 839 & 829 & 804 \\ \hline
\end{tabular}%
}
\end{table}

In addition to these parameters, the randomness of the signatures can be evaluated by measuring the entropy, which can be estimated from the data in Table \ref{random}. As described in Section \ref{threeB}, CCS challenges the $\mathrm{IED_{PUF}}$ probe with a sequence of twelve random port numbers denoted as S1, S2, ..., S12 in Table \ref{random}. About 10,000 sequences were generated to evaluate the randomness. In the table, the first row labeled D0 represents the number of occurrences of the diode D\textsubscript{0} in the sequence positions S1, S2, ..., S12 across the challenges. Similarly, the rows labeled D1, D2, and so on, represent the occurrences of diodes D\textsubscript{1}, D\textsubscript{2}, and subsequent ones. The entropy (say, $H_k$) of the sequence at the  $k$-th position can be calculated using the formula: $H_k = -\Sigma{P(D_i)}\times{\log_{2}{P(D_i)}}$, where $P(D_i)$ is the probability of occurrence of the $i$-th diode appearing in $S_k$. The maximum entropy ($H_{max}$) for any $S_k$ occurs when $P(D_0) = P(D_1) = ... P(D_{11})$, resulting in $H_{max}=3.585$. On evaluation of the entropies
for S1, S2, ..., S12 from Table \ref{random}, all entropy values were
approximately equal to $H_{max}$, with an error of less than $0.03\%$.

\section{Conclusion}\label{five}
% In this paper, a hardware-level security solution for commercial IEDs has been proposed. The solution employs physical unclonable function-based techniques to extract unique signatures from IEDs. The purpose of this work can be underlined as enhancing the security of the IEDs against counterfeiting. In this paper, the PUF probe has been presented as a complete solution with its hardware architecture and software development. The proposed PUF probe is a simple, low-cost embedded device and can be easily customized for different IED models. In this work, we have also brought out the efficacy of the proposed PUF probe with experimental results. The results also bring out the scopes for further hardening of the PUF against external variations such as power supply, ageing. Our future work will be focused in this direction and exploring signatures from IEDs of different manufacturers.
This paper introduces a hardware-level security solution specifically designed for commercial IEDs. The solution incorporates (PUF)-based techniques to extract unique signatures from the IEDs. The primary objective of this study is to enhance the security of IEDs, mitigating the risks associated with counterfeiting.
In this work, the $\mathrm{IED_{PUF}}$ probe has been presented as a complete solution, encompassing both its hardware architecture and software development. The proposed $\mathrm{IED_{PUF}}$ probe is a simple, low-cost embedded device that can be easily customized to accommodate various IED models. To validate the effectiveness of the proposed $\mathrm{IED_{PUF}}$ probe, extensive experimental evaluations have been conducted, yielding promising results.
Additionally, the experimental results highlight the scopes for further PUF hardening against external variations like power supply and aging. As part of our future work, we intend to focus on addressing these challenges and exploring the extraction of signatures from IEDs of different manufacturers.

\end{document}